\documentclass[letterpaper,12pt]{article}
\usepackage{color,amssymb,enumerate,float}
\usepackage{amsmath}

\usepackage{ifpdf}
\ifpdf
	\usepackage[pdftex]{graphicx}
	\usepackage[pdftex,unicode,implicit]{hyperref}

	\hypersetup{
  	pdftitle     = {}, 
  	pdfkeywords  = {},
  	pdfauthor    = {},
  	pdfcreator   = {pdf\LaTeXe\ with package \flqq hyperref\frqq},
  	pdfproducer  = {pdf\LaTeXe\ with package \flqq hyperref\frqq},
  	pdfpagemode  = UseNone,  
  	pdffitwindow = true,  
  	unicode      = true,
  	plainpages   = true,
  	colorlinks   = true,  
  	citecolor    = black,  
  	urlcolor     = blue, 
  	linkcolor    = black
	}

\else

  \usepackage[dvips]{graphicx}
  
\fi 

\makeatletter
\@addtoreset{equation}{section}
\makeatother

\begin{document}

\thispagestyle{empty}
	
\begin{center}
 {\bf \LARGE Hamiltonian equations of motion of quadratic gravity}
 \vspace*{15mm}
		
 {\large Jorge Bellorin}

 \vspace{3ex}
	
 {\it Department of Physics, Universidad de Antofagasta, 1240000 Antofagasta, Chile.}
		
 \vspace*{2ex}
 {\tt jorge.bellorin@uantof.cl}

 \vspace*{15mm}
 {\bf Abstract}
\begin{quotation}{\small}
We compute explicitly the equations of motion of the Hamiltonian formulation of quadratic gravity. This is the theory with the most general Lagrangian  with terms of quadratic order in the curvature tensor (discarding the cosmological constant). We employ the symbolic computational tool Cadabra. We present the linearized version of the equations of motion, performing the longitudinal-transverse decomposition. We compare the linear equations with the covariant field equations, finding that, if general-relativity terms are active, the linear Hamiltonian formulation is valid only if the perturbative spatial metric is traceless, a condition that can be freely imposed by recurring to an arbitrary function. We apply the equations of motion on homogeneous and isotropic configurations, finding explicit solutions.
\end{quotation}
\end{center}
	
\newpage
\section{Introduction}
Gravity in terms of higher-order curvature has been extensively studied, both from the classical and quantum points of view. A widely known study in this area is Stelle's renormalization proof for quadratic gravity \cite{Stelle:1976gc}. This theory has the most general Lagrangian at quadratic order in curvature, discarding topological contributions. Besides the Einstein-Hilbert term, it has the $R_{\mu\nu} R^{\mu\nu}$ and $R^2$ terms\footnote{Omitting the cosmological constant term. One of our motivations is to perform perturbative analysis. Mainly for this reason, we choose to deal with a theory that admits Minkowski spacetime as background; hence without cosmological constant.}. Despite the fact that quadratic gravity is a renormalizable theory, it possesses modes with negative norm. This feature has been an obstruction to its  physical admissibility as a quantum theory of gravitation. On the other hand, since the final theory of quantum gravity has been so difficult to find, it is worth to continue studying theories that have interesting quantum properties. For example, in Ref.~\cite{Antoniadis:1986tu}, the correctness of the propagators in the presence of unstable modes and its relationship with gauge symmetry is studied. Furthermore, counterterms might signal that higher curvature theories are an important route to explore. The (perturbative) quantum version of pure general relativity is finite at one loop \cite{tHooft:1974toh}. However, at higher loops \cite{Goroff:1985sz,Goroff:1985th}, or at one loop with couplings to matter fields \cite{tHooft:1974toh,Deser:1974cz,Deser:1974xq}, it acquires counterterms of higher order in curvature. This is a widely argued motivation to consider the quantization of theories with higher curvature. A recent study on the degrees of freedom of some models with quadratic terms in curvature can be found in Ref.~\cite{Hell:2023mph}.

Since the quadratic gravity theory contains time derivatives of higher order, its Hamiltonian formulation cannot be done by performing the standard Legendre transformation on the time derivative of the first order. A theory with higher-order time derivatives requires more initial data; hence, more canonical variables in the phase space are required. A well-known approach to treat this is the Ostrogradsky method\footnote{Ostrogradsky's method was extended to systems with constrains by Gitman et al.~\cite{Gitman}.}. The central idea is to take the several orders in time derivatives of the original coordinate as the new independent coordinates. Then, the Hamiltonian formulation can be obtained by performing the Legendre transformation on all the variables. Ostrogradsky's approach reveals generic physical instabilities on theories of higher order in time derivatives. The unphysical modes of quadratic gravity can be compared, at the classical level, to an Ostrogradskian instability.

Buchbinder and Lyahovich \cite{Buchbinder:1987vp} developed a generalization of Ostrogradsky's method and applied it to quadratic gravity\footnote{The extension of this approach to multiple dimensions was done in Ref.~\cite{Buchbinder:1991ne}.}. It has the advantage that more general variables can be used as additional coordinates, unlike Ostrogradsky's method that takes the time derivatives exclusively. This generalization is quite useful for gravity, since it can be directly implemented using the Arnowitt-Deser-Misner (ADM) formalism \cite{Arnowitt:1962hi}. A technical fact in the case of quadratic gravity is that the Hamiltonian formulation must be separated in several cases depending on the values of the coupling constants. In all cases, Buchbinder and Lyahovich presented the constraints, and the number of physical degrees of freedom. They considered the quantization, adopting a class of conditions for fixing the arbitrary functions (the gauge-fixing). The functions chosen are a subset of the components of the spatial metric. Under this condition, they obtained the integrated version of the path integral, recovering the covariant Lagrangian, with a nontrivial contribution for the quantum measure.

For two of the above mentioned cases of the coupling constants, the Hamiltonian formulation was considered previously by other authors. One of them was done by Boulware \cite{Boulware:1983yj}, and it is based on the case of the coupling constants we use in this paper, as we explain below. The constraint algebra was carefully analyzed by Boulware, and also several issues related to the quantization. The other one is the case of Weyl gravity studied by Kaku \cite{Kaku:1982xt}, who also studied the quantization of the theory in terms of the Hamiltonian formulation. The approach of Buchbinder and Lyahovich provides the proof of validity of the classical Hamiltonian formulation, based on the equivalence between the Hamiltonian equations of motion and the Lagrangian ones. Further analysis on the Hamiltonian formulation of the several cases can be found in Ref.~\cite{Kluson:2013hza}\footnote{An approach for the nonperturbative quantization of quadratic gravity based on its Hamiltonian formalism was recently focused in Ref.~\cite{Salvio:2024joi}.}. The Hamiltonian formulation for a metric-affine theory with a $R^2$ term was studied in Ref.~\cite{Glavan:2023cuy}\footnote{The Hamiltonian formulation for a bimetric gravity theory with $R^2$ terms for both metrics was studied in Ref.~\cite{Gialamas:2023fly}.}.

In this paper we present explicitly the equations of motion corresponding to the Hamiltonian formulation of quadratic gravity. The explicit form of these equations was missing in the literature. We consider that this is a significant advance in the quadratic gravity theory, since the Hamiltonian equations of motion are conceptually important on their own. They are appropriate to analyze the dynamics of the theory. In this sense, our study is the analogue of the well-known equations of motion of the ADM formulation of general relativity \cite{Arnowitt:1962hi}. The classical Hamiltonian equations are useful in many aspects. A formal question as the the correct definition of the initial value problem can be appropriately studied within this formalism. They are also useful for practical applications. For example, in numerical gravitation, the integration is widely based on field equations of first order in time derivatives. Here we present a practical analytical application: the case of homogeneous and isotropic solutions, where the field variables depend only on time. Similar studies are presented in Refs.~\cite{Demaret:1995wp,Demaret:1998dm}\footnote{The ground state, with an homogeneous and isotropic background, of a theory with a $R^2$ term was studied in Ref.~\cite{Chakraborty:2020ktp}, using the Hamiltonian formulation.}. Some studies devoted to spherically symmetric solutions of quadratic gravity can be found in Refs.~\cite{Stelle:1977ry,Lu:2015psa}. Regarding the several cases of the coupling constants, we consider only the case when the hypermatrix that arises in the relation between a time derivative and the corresponding momentum (the analogue of the de Witt metric) is invertible. This case can be considered as the generic version of the Hamiltonian formulation of quadratic gravity. To obtain the equations of motion, a previous step is to obtain the explicit form of the Poisson brackets between the independent canonical fields and the constraints. We present these brackets explicitly, including some of them that had not been shown explicitly in the literature. Related to this, we discuss about the role of the constraints as generators of transformations, in particular for the less-known case of the constraint usually associated with timelike diffeomorphisms.  We also study the linearized version of the equations of motion, where the time evolution of  several modes can be clearly characterized\footnote{In Ref.~\cite{Edery:2019bsh}, a liner-order perturbative analysis of the covariant field equations of quadratic gravity around de Sitter and Anti-de Sitter spaces was done, finding a condition of criticality on the space of coupling constants.}.

An important check of consistency is the explicit equivalence between the Hamiltonian equations of motion and the Lagrangian or covariant ones. The general method developed by Buchbinder and Lyahovich guarantees the equivalence between classical dynamics of higher orders at the level of the equations of motion. However, for a specific theory with first-class constraints, there are arbitrary functions not fixed by the equations of motion. They could play a role when establishing the equivalence between both formulations, in the sense that an arbitrary function on one side does not necessarily match with a similar function on the other side, but instead it must be fixed in order to hold the equivalence. We would like to emphasize the importance of this analysis: any loss of arbitrariness on the additional variables that could result from the check of the equivalence between both formulations -affects not only the classical theory- but also the quantum theory, since it is obtained by canonical quantization and requires the classical Hamiltonian formulation to reproduce the appropriate dynamics. We investigate this point for quadratic gravity in the context of the linearized theory.

The derivation of the Hamiltonian equations of motion of quadratic gravity is a big task. To this end, and for most of the computations in this paper, we use the computational tool Cadabra\footnote{Cadabra is found in \texttt{https://cadabra.science/}\,.} \cite{Peeters:2006kp,Peeters:2007wn,Peeters:2018dyg}. We emphasize the flexibility and easy-to-use of Cadabra to perform this kind of computations, principally due to its ability to deal with indices. The declaration of an object with upper/lower indices is just the same as writing in Latex, and contraction of repeated indices is automatic. It can handle simultaneously indices for a given space and its subspaces. This is very useful for the decomposition between time and space one performs in the Hamiltonian formulation. Commands for raising and lowering indices with metrics are incorporated, and it can deal with several metrics and covariant derivatives. We also highlight the usefulness of the \texttt{split\_index} command, which splits the sums of all the contracted indices according to the specified rule. Other powerful commands are \texttt{vary} and \texttt{integrate\_by\_parts} for the calculus of variations.
 
\section{Hamiltonian formulation}
In this section we present the classical Hamiltonian formulation of quadratic gravity developed by Buchbinder and Lyahovich \cite{Buchbinder:1987vp}. We adopt the following conventions: spacetime signature is $(-+++)$. Curvature tensors are 
\begin{equation}
 R_{\alpha\beta\mu}{}^{\nu} =
   2 \partial_{ [ \alpha } \Gamma_{ \beta ] \mu }{}^{\nu} 
 - 2 \Gamma_{ \mu [ \alpha}{}^{\rho} \Gamma_{ \beta] \rho}{}^{\nu} \,,
 \quad
 R_{\mu\nu} = R_{\alpha\mu\nu}{}^{\alpha} \,.
 \label{Ricci}
\end{equation}
To avoid confusion, we distinguish spacetime tensors with the label $^{\text{\tiny(4)}}$ and use the standard notation for spatial tensors. Thus, $R_{ij}$ and $R$ refers respectively to the spatial Ricci tensor and the spatial Ricci scalar, and so on. We assume the existence of a local coordinate system $(x^0 , x^i)$, where $x^0$ labels the time coordinate and $x^i$ the spatial coordinates. The ADM parametrization of the spacetime metric is
\begin{equation}
 g^{\text{\tiny(4)}}_{\mu\nu} = 
 \left( \begin{array}{cc}
  - N^2 + N_i N^i & N_i \\
  N_i & g_{ij}
 \end{array}\right) \,.
\end{equation}
Spatial indices are raised and lowered with $g_{ij}$.  We use the shorthands $\nabla_{ijk\cdots} = \nabla_i \nabla_j \nabla_k\cdots$, and $\nabla^2 = \nabla_k \nabla^k$ on any (density) tensor field. In the ADM frame, the extrinsic curvature tensor of the spatial hypersurfaces takes the form
\begin{equation}
	K_{ij} = \frac{1}{2N} \left( \dot{g}_{ij} - 2 \nabla_{(i} N_{j)} \right) \,.
	\label{K}
\end{equation}

The Lagrangian of quadratic gravity is \cite{Stelle:1976gc}
\begin{equation}
 \mathcal{L} = 
 \sqrt{-g^{\text{\tiny(4)}}} \left( - \kappa^{-2} R^{\text{\tiny(4)}} 
 + \alpha R^{\text{\tiny(4)}}_{\mu\nu} R^{\text{\tiny(4)}}{}^{\mu\nu}
   + \beta {R^{\text{\tiny(4)}}}^2 \right) \,,
\label{LagrangianR2}
\end{equation}
where $\kappa^{-2}$, $\alpha$ and $\beta$ are arbitrary coupling constants. We can write this Lagrangian in terms on the ADM variables, obtaining
\begin{equation}
\begin{split}
 \mathcal{L} = \,&
 \sqrt{g} N \Big[ 
 - \kappa^{-2} \left( 2 E + R + L \right)
 + \alpha \left( ( E_{ij} + R_{ij} ) ( E^{ij} +  R^{ij} ) + ( E + L )^2 
 - 2 V_i V^i \right)
\\ &
 + \beta \left( 2 E + R + L \right)^2 \Big] \,,
\end{split}
\label{LagrangianADM}
\end{equation}
where
\begin{eqnarray}
   &&
	E_{ij} = 
	\frac{1}{N} \left( \dot{K}_{ij} - \nabla_{ij} N 
	- N ( 2 K_{ik} K_{j}{}^{k} - K K_{ij} )
	- 2 K_{k(i} \nabla_{j)} N^k - N^k \nabla_k K_{ij} \right) \,,
	\\ &&
	L = K_{ij} K^{ij} - K^2 \,,
	\\ &&
	V_i = \nabla^j K_{ij} - \nabla_i K  \,,
\end{eqnarray}
and $K = g^{ij} K_{ij}$, $E = g^{ij} E_{ij}$.

The approach developed in \cite{Buchbinder:1987vp} is a modification of the approach for a Lagrangian with higher-order time derivatives developed by Ostrogradsky. The additional canonical variables are associated with the time derivatives of higher order, such that the final Hamiltonian formulation on the extended phase space is indeed of first order in time derivatives. Ostrogradsky's method requires the extended canonical variables to be exclusively of the form $q^{s + 1} = d^s x / dt^s$. The method presented in \cite{Buchbinder:1987vp} allows us to use more general functions. The canonical variables playing the role of position are $x$ and the extended variables $k^{s+1} = K^{s}(x,\dot{x},\ddot{x},...)$, $s = 1,\ldots,n-1$, where $n$ is the highest order in time derivatives of $x$ arising in the Lagrangian, and the functions $K^{s}$ depend up to the $n-1$ order of these derivatives. The requisite on the functions $K^{s}$ is that all time derivatives of $x$, up to the $n-1$ order, can be solved from the set of equalities $k^{s+1} = K^{s}(x,\dot{x},\ddot{x},...)$. The $n$-order time derivative of $x$ can be expressed as a function of the $k^{s+1}$ and the first time derivative $\dot{k}^{n}$. To obtain the Hamiltonian, the dependence of the Lagrangian on time derivatives is changed by performing Legendre transformations involving all the new variables.

The Lagrangian (\ref{LagrangianADM}) of quadratic gravity depends nonlinearly on $\dot{K}_{ij}$; hence it is a Lagrangian of second order in time derivatives of $g_{ij}$. An additional variable that is a function of the first derivative $\dot{g}_{ij}$ is required. The convenient choice is $K_{ij}$, whose expression in (\ref{K}) allows us to solve $\dot{g}_{ij} = 2 N K_{ij} + 2 \nabla_{(i} N_{j)}$. Thus, the phase space is spanned by the independent canonical pairs  $\{ (g_{ij},\pi^{ij}), (K_{ij},P^{ij}), (N,P_N), (N^i,P_i) \}$. With this definition, the unique time derivative arising in the Lagrangian is the first derivative $\dot{K}_{ij}$.

With the aim of abbreviating the notation, we introduce the index $A = 0,i\,$, such that $N$ and $N^i$ can be grouped in the set $N^A = (N,N^i)$, and similarly $P_A = (P_N, P_i)$. We list the several combinations we use in the Hamiltonian formulation:
\begin{equation}
\begin{split}
 &
 \upsilon_m = \alpha + m\beta \,, \quad  m=1,2,3\ldots,
 \\&
 \gamma = ( 8 \alpha \upsilon_3 )^{-1} \,,
 \\&
 V_i = \nabla^j K_{ij} - \nabla_i K  \,,
 \\&
 L = K_{ij} K^{ij} - K^2 \,,
 \\&
 J_{ijkl} = 2 K_{ik} K_{jl} - K_{ij} K_{kl} \,,
 \quad J_{ij} = g^{kl} J_{ijkl} \,,
 \\&
 B_{ij} = 
 \left( \nabla_{ij} + J_{ij} \right) N
 + 2 K_{k(i} \nabla_{j)} N^k + N^k \nabla_k K_{ij} 
 \,, 
 \\&  
 G^{ijkl} =  
   \alpha ( g^{ik} g^{jl} + g^{il} g^{jk} )
 + 2 \upsilon_4 g^{ij} g^{kl} \,,
 \\&
{G}^{-1}_{ijkl} =
   (4\alpha)^{-1} ( g_{ik} g_{jl} + g_{il} g_{jk} )
 - \gamma\upsilon_4 g_{ij} g_{kl} \,,
 \\&
 Q^{i j} = 
 P^{i j} - 2 \sqrt{g} \left( \alpha  R^{ij} - g^{ij} X \right)  
 \,, 
 \\& 
 \tilde{Q}_{ij} = G^{-1}_{ijkl} Q^{kl}  \,,
 \quad
 \hat{Q}_{ij} = \tilde{Q}_{ij} + \sqrt{g} R_{ij} \,,
 \\&
 V =  
 \sqrt{g} \left( 
 \kappa^{-2} ( L + R ) - \alpha ( L^2 + R_{ij} R^{ij} - 2 V_i V^i ) - \beta ( L + R )^2 \right)
 \,,
 \\&
 X = \kappa^{-2} - \upsilon_2 L - 2 \beta R \,, 
 \\& 
 Y = ( 2 \upsilon_3 )^{-1}
 \left( 
   \beta P 
 - 2 \alpha \sqrt{g} \left( 
   \kappa^{-2} 
 + \beta ( L - R ) 
 \right) \right) \,, 
 \\&
 Z = ( 2 \upsilon_3 )^{-1}
 \left(
   \upsilon_2 P
 + 2 \alpha \sqrt{g} \left(
   \kappa^{-2} 
 + \upsilon_4 ( L - R ) 
 \right) \right) \,.
\end{split}
\end{equation}
We denote traces by eliminating the indices: $K = g^{ij} K_{ij}$, $\pi = g_{ij} \pi^{ij}$, $P = g_{ij} P^{ij}$, $Q = g_{ij} Q^{ij}$ and $ \tilde{Q} = g^{ij} \tilde{Q}_{ij}$.

The canonical momentum conjugate to $K_{ij}$ is defined by
\begin{equation}
P^{ij} = \frac{ \delta \mathcal{L} }{ \delta \dot{K}_{ij} } =
\sqrt{g} G^{ijkl} E_{kl} 
+ 2 \sqrt{g} \left( \alpha  R^{ij} - g^{ij} X \right) \,. 
\label{Pij}
\end{equation}
The matrix $G^{ijkl}$ has inverse $G^{-1}_{ijkl}$: $G^{i j k l} G^{-1}_{k l m n} = \frac{1}{2} ( \delta^i_m \delta^j_n + \delta^i_n \delta^j_m )$. Throughout this analysis we assume $\alpha \neq 0$ and $\upsilon_3= \alpha + 3\beta \neq 0$, such that the inverse does exist. We consider this as the generic case of quadratic gravity. It is one the cases studied in \cite{Buchbinder:1987vp}. Velocities $\dot{K}_{ij}$ can be solved from (\ref{Pij}), yielding
\begin{equation}
\dot{K}_{ij} = 
\frac{N}{\sqrt{g}} \tilde{Q}_{i j} + B_{ij} \,.
\label{dotK}
\end{equation}
The Lagrangian (\ref{LagrangianADM}) is independent of time derivatives of $N^{A}$. This leads to the four primary constraints:
\begin{equation}
P_A = 0 \,.
\label{constrainstPA}
\end{equation}
The canonical Lagrangian is defined by 
\begin{equation}
 \mathcal{L}_{\text{CAN}} = 
 \pi^{ij} \left( 
    \dot{g}_{ij} - 2 ( N K_{ij} + \nabla_{i} N_{j} ) \right)
 + P^{ij} ( \dot{K}_{ij} - v_{ij} )|_{v_{ij}}
 + P_A  ( \dot{N}^{A} - \lambda^A ) 
 + \mathcal{L}(g_{ij},K_{ij},N^A,v_{ij})|_{v_{ij}} \,,
\end{equation}
where $\mathcal{L}(g_{ij},K_{ij},N^A,v_{ij})$ is the Lagrangian (\ref{LagrangianADM}). The variable $v_{ij}$ has been introduced to indicate that $v_{ij} = \dot{K}_{ij}$, where this $\dot{K}_{ij}$ must be replaced by its expression in terms of the canonical fields given in (\ref{dotK}). The four variables $\lambda^A$ remain as arbitrary functions, reflecting the fact that the Lagrangian (\ref{LagrangianADM}) is independent of time derivatives of $N^A$. By substituting $v_{ij}$ in $\mathcal{L}_{\text{CAN}}$, we obtain the Hamiltonian density
\begin{equation}
 \mathcal{H} =
   \frac{N}{2\sqrt{g}} \tilde{Q}_{ij} Q^{ij}
 + P^{ij} B_{ij} + 2 \pi^{ij} ( N K_{ij} + \nabla_{i} N_{j} ) 
 + N V + \lambda^A P_A \,.
\end{equation}
After some integrations by parts\footnote{Throughout this analysis, we consider only the bulk part of the functionals, assuming that all functionals are differentiable. Hence, we do not consider boundary terms explicitly.}, the Hamiltonian can be written in the form
\begin{equation}
 H = \int d^3x \left( N^A T_A + \lambda^A P_A \right) \,,
 \label{primaryHamiltomian}
\end{equation}
where
\begin{eqnarray}
 &&
 T_0 = 
   \frac{1}{2\sqrt{g}} \tilde{Q}_{ij} Q^{ij}
 + \left( \nabla_{ij} + J_{ij} \right) P^{ij} 
 + 2 K_{ij} \pi^{ij} + V \,,
 \label{T0}
 \\&& 
 T_i =
 - 2 g_{ij} \nabla_k \pi^{jk} 
 - 2 \nabla_k ( K_{ij} P^{jk} ) 
 + \nabla_i K_{jk} P^{jk}  \,.
 \label{Ti}
\end{eqnarray}

In terms of equal-time Poisson brackets, the four $P_A$ commute between them, and the four $T_A$ are independent of $N^A$. As a consequence, the preservation of the constraints (\ref{constrainstPA}) leads to the four secondary constraints:
\begin{equation}
T_A = 0 \,.
\end{equation}

\section{The algebra of constraints}
To present the algebra of the $T_A = 0$ constraints, we use a simplified notation for total spatial integrals: $\int\!F \equiv {\displaystyle \int\!d^3x F }$, where $F$ is an arbitrary function of time and space. The basic Poisson brackets of the canonical variables with $T_i$, taking an arbitrary time-dependent spatial test vector $\omega^i$, result:
\begin{equation}
\begin{split}
&
\left\{ g_{i j} \,, { \textstyle \int\! \omega^k T_k } \right\} = 
\omega^k \partial_{k} g_{i j} + 2 g_{k(i} \partial_{j)} \omega^{k} \,,
\\&
\left\{ \pi^{i j} \,, {\textstyle \int\! \omega^k } T_k \right\} = 
  \omega^k \partial_{k} \pi^{i j} 
- 2 \pi^{k(i} \partial_{k} \omega^{j)} 
+ \pi^{i j} \partial_{k} \omega^k \,,
\\&
\left\{ K_{i j} \,, {\textstyle \int \omega^k T_k } \right\} = 
\omega^k \partial_{k} K_{i j} + 2 K_{k(i} \partial_{j)} \omega^{k} \,,
\\&
\left\{ P^{i j} \,, {\textstyle \int\!\omega^k T_k } \right\} = 
  \omega^k \partial_{k} P^{i j} 
- 2 P^{k(i} \partial_{k} \omega^{j)} 
+ P^{i j} \partial_{k} \omega^k  \,.
\end{split}
\label{spatialdiff}
\end{equation}
We have then that $T_{i}$ is the generator of the spatial diffeomorphisms on these canonical variables. The brackets with the constraint $T_0$, taking an arbitrary test function $\sigma$ of the time and space, are
\begin{eqnarray}
 &&
 \left\{ g_{i j} \,, {\textstyle \int \! \sigma T_0 } \right\} = 
   2 \sigma K_{i j} \,,
 \label{bracketgT0}
 \\ && 
 \left\{ \pi^{i j} \,, { \textstyle \int \! \sigma T_0  } \right\} =  
 \frac{1}{\sqrt{g}} \left( 
   \frac{1}{4} g^{i j} Q^{kl} \tilde{Q}_{k l}  
 - \frac{1}{2\alpha} Q^{ik} Q^j{}_k
 +  \gamma\upsilon_4  Q Q^{i j}
 \right) \sigma
 \nonumber \\&&
 + \alpha \left( 
   2 \nabla^{k(i} ( \hat{Q}^{j)}{}_k \sigma ) 
 - \nabla^2 ( \hat{Q}^{ij} \sigma ) 
 - g^{i j} \nabla^{kl} ( \hat{Q}_{kl} \sigma )  
 \right)
 \nonumber\\&&
 + \left( \nabla^{ij} - g^{i j} \nabla^2 - R^{i j} \right) ( Y \sigma )
 - \nabla_{k} ( P^{k(i} \nabla^{j)}{\sigma} )
 + \frac{1}{2} \nabla_{k} ( P^{i j} \nabla^{k} \sigma  )
 \nonumber \\&&
 + \Big( 
   P^{k l} J_{kl}{}^{ij}
 + Z ( K K^{i j} - K^{ik} K^j{}_k )
 + 2 X \tilde{Q}^{ij} 
 - 4 \alpha  R^{k(i} \tilde{Q}^{j)}{}_k
 - 2 \alpha \sqrt{g} R^{ik} R^j{}_k 
 \nonumber\\&& 
 + 2 \alpha \sqrt{g} V^{i} V^{j}  
 - \frac{1}{2} g^{i j} \left( 
   2 X \tilde{Q} 
 - 2 \alpha R^{k l} \tilde{Q}_{kl} 
 + V \right)
 \Big) \sigma
 \nonumber \\&&
 + 2 \alpha \sqrt{g} \left(
   K^{i j} \nabla_{k} ( V^{k} \sigma )   
 - V^{k} \nabla_{k}{K^{i j}} \sigma 
 - 2 K^{k(i} \nabla^{j)} ( V_{k} \sigma )
 + g^{i j} \nabla_{k}( V_{l} K^{k l} \sigma ) \right)
 \,,
  \\&&
  \left\{ K_{i j} \,, {\textstyle\int \! \sigma T_0 } \right\} =   
  \frac{1}{\sqrt{g}} \tilde{Q}_{ij} \sigma
  + \left( \nabla_{i j} + J_{ij} \right) \sigma \,,
 \\&&
 \left\{ P^{i j} \,, { \textstyle \int \! \sigma T_0 } \right\} = 
 \left( 
 - 2 \pi^{i j}
 + P^{i j} K - 4 P^{k (i} K^{j)}{}_{k} + Z K^{i j}
 + g^{i j} ( P^{k l} K_{k l} - Z K ) 
 \right) \sigma 
 \nonumber  \\&&
 + 4 \alpha \sqrt{g} \left( \nabla^{(i} ( V^{j)} \sigma )
 -  g^{i j} \nabla_{k} ( V^{k} \sigma ) \right) \,.
 \label{bracketPT0}
\end{eqnarray}
The brackets (\ref{bracketgT0}) -- (\ref{bracketPT0}) were not shown explicitly in the literature previously.

On the basis of the previous brackets, we may compute the algebra the $T_A$ constraints, which takes the form
\begin{eqnarray}
 &&
 {\textstyle 
 \left\{ \int\! \omega^{i} T_{i} \,, \int\! \eta^{j} T_{j}  \right\}
 =
 \int\! \left( \omega^{i} \partial_{i} \eta^{j}
 - \eta^{i} \partial_{i} \omega^{j} \right) T_{j} } \,,
 \\&&                 
 {\textstyle 
 \left\{ \int\! \omega^{i} T_{i} \,, \int\! \sigma T_{0}  \right\}
 =
 \int\! \omega^{i} \partial_i \sigma T_0 } \,,
 \\&&
 {\textstyle 
 \left\{ \int\! \rho T_0  \,, \int\! \sigma T_{0}  \right \} =
 \int\! ( \rho \partial^i \sigma - \sigma \partial^i \rho ) T_i } \,.
\end{eqnarray}
Due to this, no more constraints are generated when the preservation of the $T_A$ constraints with the Hamiltonian (\ref{primaryHamiltomian}) is required. We have then that the full set of constraints $\{ P^A ,T_A \}$ is a set of first-class constraints (the Hamiltonian (\ref{primaryHamiltomian}) is also a first-class function). Now we may eliminate spurious variables associated with the original Lagrangian (\ref{LagrangianR2}). Constraints $P_A  = 0$ can be regarded as solved conditions, and the variables $\lambda^A$ are eliminated. The only remaining constraints are the $T_A$ constraints, which are of first class, and the fields $N^A$ play the role of their Lagrange multipliers. The phase space is spanned by the canonical pairs $\{ (g_{ij}, \pi^{ij}) , (K_{ij}, P^{ij}) \}$. The Hamiltonian is given as a sum of constraints,
\begin{equation}
 H = \int d^3x N^A T_A \,,
 \label{primaryHamiltomianfinal}
\end{equation}
such that the canonical action takes the form
\begin{equation}
	S =
	\int\! dt d^3x \left(
	\pi^{ij} \dot{g}_{ij} + P^{ij} \dot{K}_{ij} 
	- N^A T_A 
	\right) \,.
	\label{canonicalaction}
\end{equation}
The number of physical degrees of freedom in the phase space is given by (Number of canonical variables) $- 2 \times$(Number of first-class constraints) = 16. Therefore, the theory has 8 physical modes. 

\section{First-class constraints as generators}
The four first-class constraints $T_A$ of quadratic gravity satisfy exactly the same algebra of the four constraints of general relativity. This suggests that the interpretation of the transformations they generate is the same in both theories.

To study this point, let us start with the well known gauge symmetry associated with the constraint $T_i$. According to the brackets given in (\ref{spatialdiff}), this is the generator of time-dependent spatial diffeomorphisms on all the canonical variables $\psi = g_{ij}, \pi^{ij} , K_{ij}, P^{ij}$. These transformations must be complemented with the transformations of the Lagrange multipliers $N^A$, which are defined in such a way that the canonical action is left invariant. Let $\delta_{\omega^k}$ denote the complete transformation with a time-dependent spatial vector $\omega^k$ as parameter. The complete transformations are defined by
\begin{eqnarray}
 &&
 \delta_{\omega^k} \psi = 
 \left\{ \psi \,, { \textstyle \int\! \omega^k T_k } \right\} =
 \mathcal{L}_{\omega^k} \psi \,, 
 \\&&
 \delta_{\omega^k} N = 
 \mathcal{L}_{\omega^k} N = 
 \omega^k \partial_k N  \,,
 \\&&
 \delta_{\omega^k} N^i = 
 \mathcal{L}_{\omega^k} N^i + \dot{\omega}^i =
 \omega^k \partial_k N^i - N^k \partial_k \omega^i + \dot{\omega}^i \,.
\end{eqnarray}
Among these variables, only $N^i$ has a transformation that is not functionally identical to a spatial diffeomorphism along $\omega^k$, due to the term $\dot{\omega}^i$. By straightforward computations, we obtain the invariance of the sector:
\begin{equation}
\delta_{\omega^k}\! \int\! dt d^3x \left(
\pi^{ij} \dot{g}_{ij} + P^{ij} \dot{K}_{ij} 
- N^i T_i 
\right) = 0\,.
\label{deltaspatial}
\end{equation}
An intermediate step to get (\ref{deltaspatial}) is the cancellation between the time derivative $\dot{\omega}^i$ coming from the kinetic terms and the transformation of $N^i$. The invariance of the remaining sector of the action,
\begin{equation}
\delta_{\omega^k}\! \int\! dt d^3x  N T_0  = 0 \,,
\end{equation}
is automatic since this term is independent of any time derivative, and $N,T_0$ transform as scalar/density under spatial diffeomorphisms. The gauge symmetry generated by $T_i$ is an exact off-shell symmetry of the canonical action; the equations of motion are not imposed to obtain the invariance. Thus, the role of $T_i$ as generator of gauge symmetry of the canonical action is analogue to its counterpart in general relativity.

The transformation generated by $T_0$ is qualitatively different to the previous one since it does not generate diffeomorphisms when it is implemented off-shell. On the other hand, when it is implemented on-shell, that is, when the equations of motion are satisfied (and equating arbitrary parameters), it generates diffeomorphisms orthogonal to the spatial hypersurface, at least on part of the canonical variables. We may see this in the frame provided by the ADM parametrization. Let $\delta_\omega$ denote a transformation on a canonical function generated by $T_0$, with the (spatially) scalar $\omega$ as parameter,
\begin{equation}
 \delta_\omega \psi
 = 
 \left\{ \psi \,, { \textstyle \int\! \omega T_0 } \right\} \,.
\end{equation}
Since the Hamiltonian (\ref{primaryHamiltomianfinal}) is a sum of constraints, the equations of motion have the general form,
\begin{equation}
\dot{\psi} = \delta_{N} \psi + \delta_{N^i} \psi \,.
\label{EOMgeneral}
\end{equation}
We have seen that $N^i T_i$ generates a time-dependent spatial diffeomorphisms on $\psi$ along $N^i$; hence we may write
\begin{equation}
\delta_{N} \psi = \dot{\psi} - \mathcal{L}_{N^i} \psi \,.
\label{actionT0general}
\end{equation}
This relation can be interpreted as the on-shell action of the first-class constraint $T_0$. Explicitly, the transformations on the canonical fields result
\begin{eqnarray}
 &&
 \delta_{N} g_{i j} =
 \dot{g}_{ij} - N^k \partial_k g_{ij} - 2 g_{k(i} \partial_{j)} N^k \,,
 \label{diffortg}
 \\&&
 \delta_{N} K_{i j} =
 \dot{K}_{ij} - N^k \partial_k K_{ij} - 2 K_{k(i} \partial_{j)} N^k \,,
 \label{diffortK}
 \\&&
 \delta_{N} \pi^{i j} =
 \dot{\pi}^{ij} - N^k \partial_{k} \pi^{ij} - \pi^{ij} \partial_k N^k
 + 2 \pi^{k(i} \partial_k N^{j)} \,.
 \label{diffortpi}
 \\&&
 \delta_{N} P^{i j} =
 \dot{P}^{ij} - N^k \partial_{k} P^{ij} - P^{ij} \partial_k N^k
 + 2 P^{k(i} \partial_k N^{j)} \,.
 \label{diffortP}
\end{eqnarray}
We want to give a geometrical interpretation for these transformations. For the $K_{ij}$ field, one may go back to its definition (\ref{K}) as the extrinsic curvature tensor. In this scenario, $g_{ij}$ and $K_{ij}$ have standard embeddings into four-dimensional tensors. The right-hand sides of expressions (\ref{diffortg}) and (\ref{diffortK}) are equal to 4D diffeomorphisms on these objects along the orthogonal vector $N n^\mu$, which has components in the frame $(t,\vec{x})$ given by $N n^\mu = (1,-N^i)$.  Hence, at least as $g_{ij}$ and $K_{ij}$ are concerned, we have that the on-shell action of the $T_0$ constraint of quadratic gravity is to generate diffeomorphisms orthogonal to the spatial hypersurface. Therefore, the role of $T_0$ is analogue to its counterpart in general relativity. For $\pi^{ij}$ and $P^{ij}$, more analysis is required to arrive at a definitive conclusion about the geometrical significance of the transformations (\ref{diffortpi}) and (\ref{diffortP}), since the embedding of these objects into four-dimensional tensor densities seems to be a rather involved issue (a starting point could be the fact that $P^{ij}$ and $\pi^{ij}$ can be solved algebraically from the equations of motion (\ref{EOMdotK}) and (\ref{EOMdotP}), respectively)\footnote{See Ref.~\cite{Thiemann:2001gmi} for the case of general relativity.}.

\section{Hamiltonian equations of motion}
\subsection{Nonperturbative equations}
The four equations of motion can be obtained explicitly by substituting the brackets (\ref{spatialdiff}) -- (\ref{bracketPT0}) in (\ref{EOMgeneral}). Thus, the Hamiltonian equations of motion of quadratic gravity are given by
\begin{eqnarray}
 \dot{g}_{ij} &=& 
 2 N K_{i j} + 2 \nabla_{(i} N_{j)} \,,
 \label{dotg}
 \\ 
 \dot{\pi}^{ij} &=& 
 \frac{N}{\sqrt{g}} \left( 
 \frac{1}{4} g^{i j} Q^{kl} \tilde{Q}_{k l}  
 - \frac{1}{2\alpha} Q^{ik} Q^j{}_k
 +  \gamma\upsilon_4  Q Q^{i j}
 \right) 
 \nonumber \\&&
 + \alpha \left( 
   2 \nabla^{k(i} ( N \hat{Q}^{j)}{}_k ) 
 - \nabla^2 ( N \hat{Q}^{ij} ) 
 - g^{i j} \nabla^{kl} ( N \hat{Q}_{kl} )  
 \right)
 \nonumber\\&&
 + \left( \nabla^{ij} - g^{i j} \nabla^2 - R^{i j} \right) ( N Y )
 - \nabla_{k} ( P^{k(i} \nabla^{j)}N )
 + \frac{1}{2} \nabla_{k} ( P^{i j} \nabla^{k} N  )
 \nonumber \\&&
 + N \Big( 
   P^{k l} J_{kl}{}^{ij} 
 + Z ( K K^{i j} - K^{ik} K^j{}_k )
 + 2 X \tilde{Q}^{ij} 
 - 4 \alpha  R^{k(i} \tilde{Q}^{j)}{}_k 
 - 2 \alpha \sqrt{g} R^{ik} R^j{}_k
 \nonumber\\&&
 + 2 \alpha \sqrt{g} V^{i} V^{j}  
 - \frac{1}{2} g^{i j} \left( 
   2 X \tilde{Q} 
 - 2 \alpha R^{kl} \tilde{Q}_{kl} 
 + V \right)
 \Big) 
 \nonumber \\&&
 + 2 \alpha \sqrt{g} \left(
   K^{i j} \nabla_{k} ( N V^{k} )   
 - N V^{k} \nabla_{k}{K^{i j}} 
 - 2 K^{k(i} \nabla^{j)} ( N V_{k} )
 + g^{i j} \nabla_{k}( N V_{l} K^{k l} ) \right)
 \nonumber\\&&
 + N^k \nabla_{k} \pi^{i j} 
 - 2 \pi^{k(i} \nabla_{k} N^{j)} 
 + \pi^{i j} \nabla_{k} N^k
 \,,
 \label{dotpi}
 \\
 \dot{K}_{i j} &=&   
 \frac{N}{\sqrt{g}} \tilde{Q}_{ij}
 + \left( \nabla_{i j} + J_{ij} \right) N
 + N^k \nabla_{k} K_{i j} + 2 K_{k(i} \nabla_{j)} N^{k} \,,
 \label{EOMdotK}
 \\
 \dot{P}^{i j} &=& 
 N \left(
 - 2 \pi^{i j} 
 + P^{i j} K - 4 P^{k (i} K^{j)}{}_{k} + Z K^{i j}
 + g^{i j} ( P^{k l} K_{k l} - Z K ) 
 \right)  
 \nonumber  \\&&
 + 4 \alpha \sqrt{g} \left( \nabla^{(i} ( N V^{j)} )
 -  g^{i j} \nabla_{k} ( N V^{k} ) \right)
 + N^k \nabla_{k} P^{i j} 
 - 2 P^{k(i} \nabla_{k} N^{j)} 
 + P^{i j} \nabla_{k} N^k  \,.
 \nonumber\\&&
 \label{EOMdotP}
\end{eqnarray}

\subsection{The linearized equations of motion}
We analyze the perturbative equations of motion, at linear order in perturbations\footnote{We assume that appropriate spatial boundary conditions are given on the perturbative fields. In particular, we assume that the flat spatial Laplacian is an invertible operator on these fields.}. We denote the perturbative metric and lapse function by
\begin{equation}
	g_{ij} = \delta_{ij} + h_{ij} \,,
	\quad
	N = 1 + n \,.
	\label{perturbativeADM}
\end{equation}
In the Hamiltonian formalism we require to complete the definition of the background. We assume that the background values of $K_{ij}$ and $\pi^{ij}$ are equal to zero. The constraint $T_i$ (\ref{Ti}) is solved at zeroth order by these conditions. The constraint $T_0$ (\ref{T0}) indicates that the field $P^{ij}$ must acquire a nonzero background value. The zeroth-order version of this constraint yields
\begin{equation}
	\frac{1}{2} G^{-1}_{ijkl} (P^{(0)}_{kl} + 2 \kappa^{-2} \delta_{kl} ) (P^{(0)}_{ij} + 2 \kappa^{-2} \delta_{ij} )
	+ \partial_{ij} P^{(0)}_{ij} = 0 \,.
\end{equation} 
To solve it, we take
\begin{equation}
	P^{(0)}_{ij} = - 2 \kappa^{-2} \delta_{ij} 
	\quad \Rightarrow \quad
	P^{ij} = -2\kappa^{-2} \delta^{ij} + p^{ij} \,,
\end{equation}
such that $p^{ij}$ denotes the perturbation of $P^{ij}$. For the rest of fields and combinations of fields that get no zeroth mode, like the shift vector $N^i$, we keep the original notation. All the equations of motion (\ref{dotg}) -- (\ref{EOMdotP}) are solved by the background as we have defined it. Three additional combinations of coupling constants are useful:
\begin{equation}
\eta = 3\alpha + 8\beta \,,
\quad
\tilde{\eta} = 5\alpha + 16\beta \,,
\quad
\mu = \frac{\kappa^{-2}}{\alpha} \,.
\end{equation}

The linear-order version of the $T_A$ constraints is
\begin{eqnarray}
	&&
	T_0 = 
	\partial_{kl} p^{kl} - \kappa^{-2} \partial_{kl} h_{kl} \,,	
	\label{lineart0}
	\\&&
	T_i = - 2 \partial_k \pi^{ki}
	+ 2 \kappa^{-2} \left( 2 \partial_k K_{ki} 
	- \partial_i K_{kk} \right) \,.
	\label{linearti}
\end{eqnarray}
The linear-order version of the equations of motion (\ref{dotg}) -- (\ref{EOMdotP}) yields
\begin{eqnarray}
 \dot{h}_{ij} &=& 
 2 K_{i j} + 2 \partial_{(i} N_{j)} \,,
 \label{lineardotg}
 \\ 
 \dot{\pi}_{ij} &=&  
 \alpha \left( 
   2 \partial_{k(i} \hat{Q}_{j)k}  
 - \Delta \hat{Q}_{ij} 
 - \delta_{ij} \partial_{kl} \hat{Q}_{kl} \right)
 - 4\alpha\gamma \tau_{ij} \Delta
 \left( \beta p^{kk} + 2\alpha\beta R - \kappa^{-2} \upsilon_2 h_{kk} 
 \right) 
 \nonumber \\&&
 + \kappa^{-2} \left(     
   2 \tilde{Q}_{ij} - \alpha \gamma \delta_{ij} Q^{kk} 
 + R_{i j} - \frac{1}{2}\delta_{ij} R 
 + \partial_{ij} n \right) 	
 \,,
 \\
 \dot{K}_{i j} &=&   
 \tilde{Q}_{ij} + \partial_{ij} n 
 \,,
 \\
 \dot{p}^{ij} &=& 
 - 2 \pi^{ij}
 + 2 \kappa^{-2} \left( 3 K_{ij}  
 + 2 \partial_{(i} N^{j)}
 - \delta_{ij} ( K_{kk} + \partial_{k} N^k ) \right)
 \nonumber \\&&
 + 4 \alpha \left( 
 \partial_{k(i} K_{j)k} - \partial_{ij} K_{kk}
 + \delta_{ij} \tau_{kl} \Delta K_{kl} \right) 
 \,,
 \label{lineardotp}
\end{eqnarray}
where $\Delta \equiv \partial_{kk}$, $\tau_{ij}$ is the transverse projector
\begin{equation}
 \tau_{ij} = \delta_{ij} - \partial_{ij} \Delta^{-1} \,, 
\end{equation}
$R_{ij}$ and $R$ are expanded up to linear order, 
\begin{eqnarray}
 &&
 Q^{kk} = 
 p^{kk} - 2 \upsilon_6 R + \kappa^{-2} h_{kk} \,, 
 \\ &&
 \hat{Q}_{ij} = 
  \frac{1}{2\alpha} p^{ij}
 - \mu h_{ij}
 - \gamma \delta_{ij} \left( 
   \upsilon_4 p^{kk} 
 - 2 \alpha \upsilon_2 R 
 - \kappa^{-2}\eta h_{kk} 
 \right) \,,
\end{eqnarray}
and $ \tilde{Q}_{ij} =  \hat{Q}_{ij} - R_{ij} $.
 
To characterize the dynamics of the linear degrees of freedom we employ the three-dimensional longitudinal-transverse decomposition: if $\sigma_{ij}$ is a symmetric spatial tensor, we decompose it in the way
\begin{equation}
		\sigma_{ij} = 
		\partial_{ij} \Delta^{-1} \sigma^L
		+ 2 \partial_{(i} \sigma_{j)}^L
		+ \frac{1}{2} \tau_{ij} \sigma^T  
		+ \sigma_{ij}^{T} \,,
	\label{decomposition}
\end{equation}
with the conditions: $\sigma_{kk}^T = \partial_k \sigma_{kj}^T = \partial_k \sigma_k^L = 0$. For the shift vector we use
\begin{equation}
	N^i = \partial_i \Delta^{-1} N^L + N_i^T \,, \quad \partial_k N_k^T = 0 \,.
	\label{decomposeNi}
\end{equation}
The decomposition of the constraint (\ref{lineart0}) yields the condition
\begin{equation}
 p^L = \kappa^{-2} h^L \,,
 \label{constpl}
\end{equation}
and the constraint (\ref{linearti}) gives us two conditions:
\begin{eqnarray}
&&
 \pi^L = \kappa^{-2} ( K^L - K^T ) \,,
 \\&&
 \pi^L_i = 2 \kappa^{-2} K^L_i \,.
 \label{constpil}
\end{eqnarray}
We move to the equations of motion. It is not necessary to consider the equations of motion of the longitudinal variables $p^L, \pi^L , \pi^L_i$ independently, since the constraints (\ref{constpl}) -- (\ref{constpil}) are preserved by the equations of motion. For the rest of canonical variables, the decomposition of the equations (\ref{lineardotg}) -- (\ref{lineardotp}), after using the solutions of the constraints (\ref{constpl}) -- (\ref{constpil}), gives us the set of equations of motion:
\begin{eqnarray}
&&
\dot{h}^L =  2 K^L + 2 N^L  
\,,	
\label{dothl}
\\&&
\dot{h}^T = 2 K^T \,,
\label{}
\\&&
\dot{K}^L = 
\gamma \left( 
- 2\kappa^{-2}\upsilon_4 h^L
+ \left( 2\alpha \upsilon_4 \Delta 
+ \kappa^{-2}\eta \right) h^T 
- \upsilon_4 p^T
\right) + \Delta n \,,
\label{}
\\&&
\dot{K}^T = 
\gamma \left(
\left( 4\alpha\beta \Delta 
- 2\kappa^{-2} \upsilon_4 \right) h^T 
+ 2\upsilon_2 p^T 
+ 4\kappa^{-2} \upsilon_2 h^L
\right) \,,
\\&&
\dot{\pi}^T =
\gamma \left( 
- ( 4\alpha\beta \Delta - 2\kappa^{-2} \upsilon_4 ) p^T
+ \left( 8\alpha^2\upsilon_4 \Delta^2	
+ 4\kappa^{-2} \alpha\upsilon_1 \Delta  
- 2\kappa^{-4}\eta \right) h^T 
\right.\nonumber\\ && \quad\quad \left.
+ ( 8 \kappa^{-2}\alpha\upsilon_2 \Delta 
+ 4\kappa^{-4}\upsilon_4 ) h^L
\right) 
\,,
\\&&
\dot{p}^T =
- 2\pi^T 
+ ( 8\alpha \Delta + 2\kappa^{-2} ) K^T 
- 4\kappa^{-2} K^L
- 4\kappa^{-2} N^L \,,
\label{dotpT}
\\&&
\dot{h}^L_i = 2 K^L_i + N_i^T \,,
\\&&
\dot{K}^L_i = 
\frac{1}{2\alpha} p^L_i - \mu h^L_i 
\,,
\\&&
\dot{p}^L_i =
\left( 2\alpha \Delta + 2\kappa^{-2} \right) K^L_i 
+ 2 \kappa^{-2} N^T_i  \,,
\label{dotpLi}
\\&&
\dot{h}^T_{ij} = 2 K_{ij}^T \,,
\label{dothtij}
\\&&
\dot{K}^T_{ij} = 
  \left( \frac{1}{2} \Delta 
- \mu \right) h^T_{ij}
+ \frac{1}{2\alpha} p^T_{ij} 
\,,
\\&&
\dot{\pi}_{ij}^T = 
- \left( \frac{1}{2} \Delta - \mu \right) p^T_{ij}
+ \kappa^{-2} \left(   
  \frac{3}{2} \Delta - 2\mu \right) h^T_{ij}
\,,
\\&&
\dot{p}_{ij}^T = 
- 2\pi_{ij}^T + 6\kappa^{-2} K^T_{ij}
\,.
\label{dotptij}
\end{eqnarray}
In the set of equations (\ref{dothl}) -- (\ref{dotptij}), there are 4 arbitrary functions of time. After subtracting them, the system determines the time evolution of 16 independent initial data given on the spatial hypersurface corresponding to the initial time. Specifically, the equations for $\dot{h}^L$, $\dot{h}^L_i$, $\dot{K}^L$, $\dot{p}^T$ and  $\dot{p}^L_i$ depend on the Lagrange multipliers $n, N^L, N^T_i$. Hence, the time evolution of 4 combinations of these canonical variables cannot be uniquely determined by the initial data. If one fixes the functions $n, N^L, N^T_i$ for all time, then the time evolution of all the canonical variables is uniquely given by the initial data. If one fixes the four combinations of canonical variables for all time, then the equations giving the time derivative of these combinations determine the Lagrange multipliers $n, N^L, N^T_i$. Since the dependence of the Eqs.~(\ref{dotpT}) and (\ref{dotpLi}) on the Lagrange multipliers is proportional to $\kappa^{-2}$, it seems natural or intrinsic to the quadratic theory to choose the 4 longitudinal canonical functions $h^L, K^L, h^L_i$ as the variables affected by the indetermination on the time evolution. 

In the next section, we discuss that the validity of the Hamiltonian formulation requires that the variable $h^L$ must be fixed by the condition $h^L = - h^T$. We take in advance this result, and reduce the previous Hamiltonian equations for the scalar modes. The Lagrange multiplier $N^L$ is fixed by the Eq.~(\ref{dothl}), such that $2 N^L = - ( \dot{h}^T + 2 K^L )$. The final set of equations of motion for all modes result:
\begin{eqnarray}
&&
\dot{h}^T = 2 K^T \,,
\label{}
\\&&
\dot{K}^L = 
\gamma \left( 
  \left( 2\alpha \upsilon_4 \Delta 
+ \kappa^{-2} \tilde{\eta} \right) h^T 
- \upsilon_4 p^T
\right) + \Delta n \,,
\label{}
\\&&
\dot{K}^T = 
\gamma \left(
\left( 4\alpha\beta \Delta 
- 2\kappa^{-2} \eta \right) h^T 
+ 2\upsilon_2 p^T 
\right) \,,
\\&&
\dot{\pi}^T =
\gamma \left( 
- ( 4\alpha\beta \Delta - 2\kappa^{-2} \upsilon_4 ) p^T
+ \left( 8\alpha^2\upsilon_4 \Delta^2	
- 4\kappa^{-2} \alpha\upsilon_3 \Delta  
- 2\kappa^{-4} \tilde{\eta} \right) h^T 
\right) \,,
\\&&
\dot{p}^T - 2\kappa^{-2} \dot{h}^T =
- 2\pi^T 
+ ( 8\alpha \Delta + 2\kappa^{-2} ) K^T  \,,
\label{}
\\&&
\dot{h}^L_i = 2 K^L_i + N_i^T \,,
\\&&
\dot{K}^L_i = 
\frac{1}{2\alpha} p^L_i - \mu h^L_i 
\,,
\\&&
\dot{p}^L_i =
\left( 2\alpha \Delta + 2\kappa^{-2} \right) K^L_i 
+ 2 \kappa^{-2} N^T_i  \,,
\label{}
\\&&
\dot{h}^T_{ij} = 2 K_{ij}^T \,,
\label{}
\\&&
\dot{K}^T_{ij} = 
\left( \frac{1}{2} \Delta 
- \mu \right) h^T_{ij}
+ \frac{1}{2\alpha} p^T_{ij} 
\,,
\\&&
\dot{\pi}_{ij}^T = 
- \left( \frac{1}{2} \Delta - \mu \right) p^T_{ij}
+ \kappa^{-2} \left(   
\frac{3}{2} \Delta - 2\mu \right) h^T_{ij}
\,,
\\&&
\dot{p}_{ij}^T = 
- 2\pi_{ij}^T + 6\kappa^{-2} K^T_{ij}
\,.
\label{}
\end{eqnarray}

\section{Covariant field equations}
The ADM variables are useful also for the covariant field equations, which are derived from the Lagrangian (\ref{LagrangianR2}). These equations take the form
\begin{equation}
\begin{split}
	&
	H_{\mu\nu} \equiv 
	\kappa^{-2} R^{\text{\tiny(4)}}_{\mu \nu} 
	- 2 \alpha R^{\text{\tiny(4)}}_{\alpha \mu} 
	  R^{\text{\tiny(4)}}{}^{\alpha}{}_{\nu} 	 
	- 2 \beta R^{\text{\tiny(4)}} R^{\text{\tiny(4)}}_{\mu \nu} 
	+ 2 \alpha \nabla^{\text{\tiny(4)}}_{\alpha(\mu}     
	  R^{\text{\tiny(4)}}_{\nu)}{}^\alpha 
	- \alpha \nabla^{\text{\tiny(4)}}_{\alpha}{}^{\alpha} 
  	  R^{\text{\tiny(4)}}_{\mu \nu} 
	+ 2 \beta \nabla^{\text{\tiny(4)}}_{\mu \nu} R^{\text{\tiny(4)}}
	\\&
	+ \frac{1}{2} g^{\text{\tiny(4)}}_{\mu \nu} \left( 
	- {\kappa}^{-2} R^{\text{\tiny(4)}}    
	+ \alpha R^{\text{\tiny(4)}}_{\alpha\beta} 
      R^{\text{\tiny(4)}}{}^{\alpha\beta}
	+ \beta {R^{\text{\tiny(4)}}}^{2}
	- \upsilon_4 \nabla^{\text{\tiny(4)}}_\alpha{}^\alpha 
	  R^{\text{\tiny(4)}} \right)  
	= 0 \,.
\end{split}
\label{covEOM}
\end{equation}
On the linearized version of the Eq.~(\ref{covEOM}), we substitute the ADM fields and then perform the longitudinal-transverse decomposition (\ref{decomposition}) -- (\ref{decomposeNi}). This procedure yields four independent field equations:
\begin{eqnarray}
 &&
   \left( \upsilon_4 \Box + \kappa^{-2} \right) h^T
 - 2 \upsilon_2 \phi = 0 \,,
 \label{covht2nd}
 \\ &&
   \left( \eta \Box + \kappa^{-2} \right) \Box h^T
 - 2 \left( \upsilon_4 \Box + \kappa^{-2} \right) \phi = 0 \,,
 \label{covht4th}
 \\&&
 \left( \Box - \mu \right) K^L_i  = 0 \,,
 \label{covkl}
 \\&&
 \left( \Box - \mu \right) \Box h^T_{ij} = 0 \,,
 \label{covhijT}
\end{eqnarray}
where
\begin{equation}
 \Box = -\partial_{00} + \Delta \,,
 \quad
 \phi = \ddot{h}^L - 2 \dot{N}^L - 2\Delta n \,,
 \quad
 K^L_i = \frac{1}{2} ( \dot{h}^L_i - N^T_i ) \,.
\end{equation}
We comment that one may obtain decoupled field equations for $\phi$ and $h^T$ in local form, by combining the Eqs.~(\ref{covht2nd}) and (\ref{covht4th}). By doing this, we obtain that $h^T$ and $\phi$ satisfy the same fourth-order decoupled equation, which is\footnote{In the case $\upsilon_2 = \alpha + 2\beta = 0$, the Eqs.~(\ref{covht2nd}) and (\ref{covht4th}) yield the second-order equation \[ \left( \Box - \mu \right) \psi = 0 \,. 
\]}
\begin{equation}
	\left( \Box + \frac{\kappa^{-2}}{2\upsilon_3} \right) 
	\left( \Box - \mu \right) \psi  = 0 \,,
	\label{covhT4}
\end{equation}
with $\psi = h^T , \phi$. The locality of Eq. (\ref{covhT4}) leads to time derivatives on $h^L$ of sixth order. The Eqs. (\ref{covkl}), (\ref{covhijT}) and (\ref{covhT4}) are written in terms of Klein-Gordon operators. They are valid in the frame defined by the ADM parametrization. Furthermore, these equations lead to dispersion relations with real solutions for the frequency if one imposes the following bounds on the coupling constants:
\begin{equation}
	\kappa^{-2} \geq 0 \,,
	\quad
	\alpha > 0 \,,
	\quad
	\beta < - \frac{1}{3} \alpha \,.
\end{equation} 
The first two conditions lead to $\mu \geq 0$ and the last one to $\upsilon_3 < 0$. We remark that, to arrive at the Eqs.~(\ref{covht2nd}) -- (\ref{covhT4}), we have not fixed the gauge.

We move to the equivalence between the Hamiltonian and covariant field equations. It turns out that the general-relativity terms induce a discontinuity in the way the arbitrary functions are handled in order to get the equivalence. First, we find that, in the $\kappa^{-2}=0$ case, both kinds of field equations are completely equivalent without fixing any of the arbitrary functions. Explicitly, when $\kappa^{-2} = 0$, the decoupled equations for the orthogonal modes obtained from (\ref{covht2nd}) -- (\ref{covhijT}) become wave and square-wave equations:
\begin{eqnarray}
	&&
	\Box \phi = 0 \,,
	\label{wavephi}
	\\&&
	\Box^2 h^T = 0 \,,
	\\&&
	\Box K^L_i  = 0 \,,
	\\&&
	\Box^2 h^T_{ij} = 0 \,.
	\label{wavehT}
\end{eqnarray} 
These equations can be obtained directly from the Hamiltonian Eqs.~(\ref{dothl}) -- (\ref{dotptij}). On the other hand, when  the general-relativity terms are active, $\kappa^{-2} \neq 0$, we find that it is necessary to fix part of the arbitrary functions to obtain the full equivalence between the field equations. Specifically, the Eqs.~(\ref{covht2nd}), (\ref{covkl}) and (\ref{covhijT}) are exactly reproduced from (\ref{dothl}) -- (\ref{dotptij}) without any additional condition. However, from the Hamiltonian equations, we obtain
\begin{equation}
  \left( \eta \Box + \kappa^{-2} \right) \Box h^T
 - 2 \left( \upsilon_4 \Box + \kappa^{-2} \right) \phi 
 = 
 - 2\kappa^{-2} \Delta ( h^L + h^T ) \,.
\end{equation}
This result leads us to take $h^L$ as one of the four arbitrary functions of the Hamiltonian formulation, and impose
\begin{equation}
 h^L = - h^T \,.
 \label{fixhL}
\end{equation}
This condition is the same as demanding that the perturbation of the spatial metric is traceless. With this, the Eq.~(\ref{covht4th}) is reproduced and both sets of field equations become equivalent. The pair of Eqs.~(\ref{covht2nd}) and (\ref{covht4th}) then determines $h^T$ and the combination $\dot{N}^L + \Delta n$. We remark that it is the equivalence between the field equations that gives support to the Hamiltonian formulation. Hence, the condition (\ref{fixhL}) is a requisite for the validity of the linearized Hamiltonian formulation of quadratic gravity.

\section{Homogeneous and isotropic solutions}
As an application of the Hamiltonian equations of motion and constraints, we consider an homogeneous and isotropic configuration. For a situation like this, the Hamiltonian equations are useful since they have time derivatives in one single term and separated from spatial derivatives. Moreover the constraints have no time derivatives. 

We impose the condition of proper time, $N = 1$, such that the homogeneous and isotropic spacetime metric can be written as
\begin{equation}
	ds^2 = - dt^2 + a(t)^2 dx^i dx^i \,.
	\label{homiso}
\end{equation}
For the simplicity of the discussion, we have considered only the case of the flat spatial geometry. For the external source we choose a perfect fluid, which is suitable for this kind of geometry. We consider the energy-momentum tensor of the form
\begin{equation}
	T_{\mu\nu} = \rho u_\mu u_\nu + P ( g^{\text{\tiny(4)}}_{\mu\nu} + u_\mu u_\nu ) \,, 
	\label{energymomentum}
\end{equation}
where the fluid is at rest in the given reference frame: $u^\mu = ( 1,0,0,0 )$. Problems of finite and infinite actions of quadratic gravity in homogeneous and isotropic background coupled to a perfect fluid have been recently studied in Ref.~\cite{Lehners:2023fud}.

To obtain a consistent coupling in the Hamiltonian formalism, we do the same as in general relativity for this class of configurations. At the end, we compare with the covariant field equations of quadratic gravity. Meanwhile, this provides another check of the validity of the Hamiltonian formulation\footnote{Notice that this case is different to the perturbative analysis of section 5.2, due to the boundary conditions. In the homogeneous and isotropic ansatz, we do not fix a Dirichlet problem on the spatial directions, since $a$ is a function only of time. Therefore, the conditions (if any) on the underlying arbitrary functions are not the same in both approaches.}. The source only arises in the $T_0$ constraint and in the $\dot{\pi}^{ij}$ equation of motion. These equations become 
\begin{eqnarray}
 &&
 T_0 = 2 a^3 \rho,
 \label{T0sourced}
 \\&&
 \dot{\pi}^{ij} = (\cdots) - a P \delta_{ij} \,,
 \label{dotpisourced}
\end{eqnarray}
where $T_0$ is given in (\ref{T0}), and the ellipsis stands for all the terms in the right-hand side of the Eq.~(\ref{dotpi}). The constraint $T_i$, given in (\ref{Ti}), and the equations of motion (\ref{dotg}), (\ref{EOMdotK}) and (\ref{EOMdotP}) maintain their source-free form.

After adopting the homogeneous and isotropic geometry (\ref{homiso}), we use the Eq.~(\ref{dotg}) to solve $K_{ij}$, Eq.~(\ref{EOMdotK}) to solve $P_{ij}$, and Eq.~(\ref{EOMdotP}) to solve $\pi^{ij}$, obtaining:
\begin{eqnarray}
 &&
 K_{ij} = a \dot{a} \,\delta_{ij} \,,
 \label{homisoK}
 \\&&
 \pi^{ij} =
 a \left(
 - 4\upsilon_3 \left(
     \frac{\dddot{a}}{a} + 2 \frac{\dot{a}}{a} \frac{\ddot{a}}{a} \right)
 + 6\upsilon_2 \left( \frac{\dot{a}}{a} \right)^3 
 + \kappa^{-2} \frac{\dot{a}}{a}
 \right) \delta_{ij} \,,
 \label{homisopi}
 \\&&
 P^{ij} = 
 2 a \left( 
 4\upsilon_3 \frac{\ddot{a}}{a}
 + 2\upsilon_6 \left( \frac{\dot{a}}{a} \right)^2
 - \kappa^{-2}
 \right) \delta_{ij} \,.
 \label{homisoP} 
\end{eqnarray}
These expressions, together with (\ref{homiso}), determine the homogeneous and isotropic configuration in the Hamiltonian formalism. The constraint $T_i = 0$ is solved automatically by these fields since all terms of this constraint contain spatial derivatives. Equivalently, the components $H_{0i}=0$ of the covariant field equations are solved by the homogeneous and isotropic configurations. After substituting the homogeneous and isotropic fields in the constraint (\ref{T0sourced}), we obtain that it acquires a rather short form,
\begin{equation}
   6\upsilon_3 \left[
 - 2 \frac{\dot{a}}{a} \frac{\dddot{a}}{a}
 + \left(\frac{\ddot{a}}{a}\right)^2
 - 2 \frac{\ddot{a}}{a} \left(\frac{\dot{a}}{a}\right)^2 
 + 3 \left(\frac{\dot{a}}{a}\right)^4
 \right]
 + 3 \kappa^{-2} \left(\frac{\dot{a}}{a}\right)^2
 = \rho  \,.
 \label{homisoT0}
\end{equation}
We have checked that the component $H_{00} = \rho$ of the covariant field equations reproduces the Eq.~(\ref{homisoT0}). Finally, all terms of the equation of motion (\ref{dotpisourced}) become proportional to $\delta_{ij}$; hence, we must consider one single equation. It also takes a short form,
\begin{equation}
  2\upsilon_3 \left[
   2 \frac{\ddddot{a}}{a}
 + 4 \frac{\dot{a}}{a} \frac{\dddot{a}}{a} 
 + 3 \left( \frac{\ddot{a}}{a} \right)^2
 - 12 \frac{\ddot{a}}{a} \left( \frac{\dot{a}}{a} \right)^2
 + 3 \left( \frac{\dot{a}}{a} \right)^4
 \right]
 - \kappa^{-2} \left[ 
     2 \frac{\ddot{a}}{a}
   + \left( \frac{\dot{a}}{a} \right)^2
 \right]
 = P \,.
 \label{homisodotpi}
\end{equation}
Again, we have checked that the components $H_{ij} = a^2 P \delta_{ij}$ of the covariant field equations are equal to the Eq.~(\ref{homisodotpi})\footnote{Notice that in the Eqs.~(\ref{homisoT0}) and (\ref{homisodotpi}), the terms of higher order in derivatives group themselves in such a way that $\upsilon_3$ is their common coupling constant. Hence, if we go to the limit $\upsilon_3 = 0$, all these terms drop out and we recover the Friedman equations. Although this a feature for this kind of configurations, the limit $\upsilon_3 = 0$ is not valid in our analysis, since we have assumed $\alpha \neq 0$ and $\upsilon_3 \neq 0$ to build the whole Hamiltonian formulation, as we have indicated in section 2.}.

As usual, we assume that there is a equation of state relating $\rho$ and $P$ in a linear form, $P = k \rho$ , where $k$ is a constant. In the case $\kappa^{-2} = 0$, we may find the solutions of the Eqs.~(\ref{homisoT0}) and (\ref{homisodotpi}), with the function $a(t)$ having the form of a  power of $t$. The first solution is
\begin{equation}
 a = C t^{4/(3(1+k))}  \,, 
 \quad 
 \rho = \frac{ 32\upsilon_3 ( - 3k^2 + 2k + 5 )}
              { 3 ( 1 + k )^4 } t^{-4} \,,
\end{equation}
where $C$ is an arbitrary constant. This solution satisfies the relation of proportionality $\rho \propto a^{-3(1+k)}$, which agrees with the continuity equation of the perfect fluid. Notice that if $k=1/3$ (radiation), the function $a$ grows linearly with time. In the case $k=0$ (dust), $a\sim t^{4/3}$. In both situations the condition $\upsilon_3 = \alpha + 3\beta > 0$ must be assumed in order to get a positive $\rho$. 
The second solution has zero matter density $\rho = 0$, and $a$ proportional to $\sqrt{t}$. Hence, it is a dynamical vacuum solution. The last solution is just Minkowski spacetime in vacuum.

\section{Conclusions}
We have presented explicitly the Hamiltonian equations of motion of quadratic gravity. These equations follow after computing the basic Poisson brackets between the canonical field variables and the constraints. We have considered only the case when the hypermatrix relating the velocities with the canonical momenta, $G^{ijkl}$, is invertible, adopting this case as the generic one. In this case no more constraints than the four $T_A = 0$ are generated, and they are first-class constraints. Throughout this analysis, we have used the computational tool Cadabra, finding that it fits very well to the kind of computations required, with a very easy and intuitive way of writing. The principal feature of Cadabra that we want to highlight is its ability to deal with objects with indices. We hope that our results can serve as one example of practical utility of this tool.

We have presented the linearized version of the Hamiltonian equations of motion, performing the longitudinal-transverse decomposition of the spatial tensors. Since the decomposition is orthogonal, each mode gets is own equation of motion. We have performed this and the nonperturbative analysis without fixing the arbitrary functions that are associated with the first-class constraints. The perturbative analysis has allowed us to perform an explicit check of the equivalence between the Hamiltonian equations of motion and the covariant ones. We have found that, when general-relativity terms are active, it is necessary to eliminate the longitudinal scalar mode $h^L$ in a specific way, which is equivalent to demand that the perturbative spatial metric is traceless. We emphasize that this condition is a requisite for the validity of the Hamiltonian formulation, at linear order in perturbations. More analysis is required in order to establish the explicit equivalence between both formulations at the level of the nonperturbative equations. This may lead to elucidate what are the conditions required on the arbitrary functions in the nonperturbative case.
 
Our study provides a missing part of the quadratic gravity theory, since the explicit form of the Hamiltonian equations of motion was not known previously. These equations are relevant on their own; they help to extract the physical features of the theory from a perspective different from the usual covariant formulation. We expect that the explicit form of the equations of motion, both the nonperturbative and perturbative versions, could be useful in different instances. As an example, we have studied the case of homogeneous and isotropic configurations coupled to a perfect fluid. We have obtained explicit solutions that exhibit consistent physical features. 

The invertibility of $G^{ijkl}$ requires in particular that the coupling constant of the $R_{\mu\nu} R^{\mu\nu}$ term cannot be turned off. As a consequence, the Hamiltonian formulation of this case is not continuously connected to the Hamiltonian formulation of general relativity; that is, one cannot obtain the ADM Hamiltonian of general relativity as a smooth limit of the case we have considered here. 
Other cases of values of the coupling constants can be considered as well. Besides the issue of the smooth connection to general relativity, there is an interesting case where the theory acquires a conformal invariance ($\upsilon_3 = 0$), which is the widely known case of Weyl gravity.

\section*{Conflict of interest}
The author states explicitly that there are no conflicts of interest in connection with this article.

\section*{Data Availability statement}
No data has been generated or used for this work.


\end{document}